\documentclass[twocolumn,aps,pre,floatfix,superscriptaddress]{revtex4}
\usepackage{amsmath,amssymb,eucal,graphicx,bm}
\begin{document}
\title{Jamming and Tiling in Aggregation of Rectangles}
\author{D.~S.~Ben-Naim}
\affiliation{Department of Computer Science, University of California
  Santa Barbara, Santa Barbara, California 93106, USA}
\author{E.~Ben-Naim}
\affiliation{Theoretical Division and Center for Nonlinear Studies,
  Los Alamos National Laboratory, Los Alamos, New Mexico 87545, USA}
\author{P.~L.~Krapivsky}
\affiliation{Department of Physics, Boston University, Boston,
  Massachusetts 02215, USA}
\begin{abstract} 
We study a random aggregation process involving rectangular clusters.
In each aggregation event, two rectangles are chosen at random and if
they have a compatible side, either vertical or horizontal, they merge
along that side to form a larger rectangle.  Starting with $N$
identical squares, this elementary event is repeated until the system
reaches a jammed state where each rectangle has two unique sides.  The
average number of frozen rectangles scales as $N^\alpha$ in the
large-$N$ limit.  The growth exponent $\alpha=0.229\pm 0.002$
characterizes statistical properties of the jammed state and the
time-dependent evolution.  We also study an aggregation process where
rectangles are embedded in a plane and interact only with nearest
neighbors. In the jammed state, neighboring rectangles are
incompatible, and these frozen rectangles form a tiling of the 
two-dimensional domain. In this case, the final number of rectangles
scales linearly with system size.
\end{abstract}
\maketitle

\section{Introduction}

Aggregation, the random process by which clusters merge irreversibly
to form larger clusters \cite{mvs16,mvs17,fl,krb}, underlies natural
and physical phenomena such as polymerization \cite{pjf, whs, pf},
evolution of accretion disks \cite{dt,sc,bkbshss}, formation of rain
and dust clouds \cite{skf,ffs,gw,pw}, as well as growth of complex
networks \cite{bb,jlr,dm,mejn}. Non-equilibrium in nature, aggregation
processes exhibit rich phenomenology including self-similar growth
\cite{fl}, condensation \cite{kr96,mkb,rm}, gelation
\cite{zhe,bj,aal,bk05} and instant gelation
\cite{dom,hez,spouge,van87,bk,pl,mg,bk03}.

The time-dependent evolution of an aggregation process is stochastic
and may exhibit significant fluctuations from one realization to
another, phase transitions, and mass distributions that sensitively
depend on microscopic details of the merger mechanism
\cite{fl,krb}. However, the final state is usually deterministic as
the system condenses into a single cluster that contains all of the
mass. Nontrivial final states with multiple clusters require an
additional competing process, for example, fragmentation where large
cluster may break into smaller ones \cite{mkb,rm,bkbshss,bk08}.

In this paper, we introduce an irreversible aggregation process that
ends in a nontrivial jammed state with multiple frozen clusters. This
final state is stochastic as the number of final clusters fluctuates
from one realization to another.  This behavior occurs because
aggregates have specific shapes which are preserved throughout the
aggregation process, thereby constraining merger events.

\begin{figure}[t]
\vspace{.4in}
\includegraphics[width=0.4\textwidth]{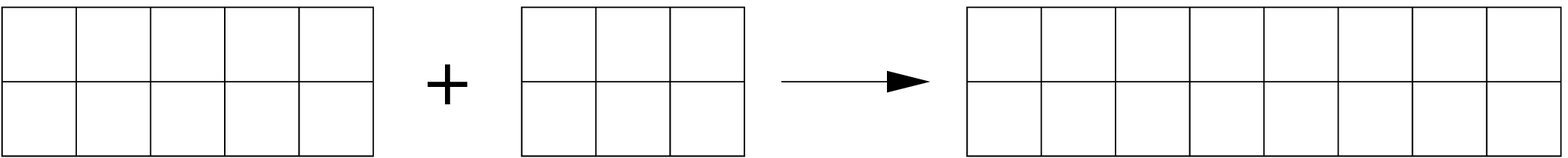}
\caption{Illustration of the aggregation process \eqref{process-ver}.}
\label{fig-agg}
\end{figure}

We study an aggregation process involving rectangles, where in each
aggregation event, two rectangles combine to form a larger rectangle. 
In each elementary step, two rectangles are chosen at random, and in
addition, one side, either the vertical one or the horizontal one, is
chosen at random.  If the two rectangles are compatible along the
randomly-chosen side, they merge along that side (figure \ref{fig-agg}
illustrates merger along the vertical side).

This aggregation process conserves area and preserves shape as all
clusters remain rectangular. Initially the system consists of $N$
identical squares. The elementary step is repeated until the system
reaches a jammed state where each rectangle has two unique sides and
aggregation is no longer feasible. We study statistical properties of
the jammed state and the time-dependent evolution toward that state.

Generally, the number of rectangles in the jammed state fluctuates
from one realization of the random aggregation process to another. Our
main result is that the average number of frozen rectangles, $F$,
grows sub-linearly with system size (figure \ref{fig-FN})
\begin{equation}
\label{FN}
F\sim N^\alpha,\quad {\rm with}\quad \alpha=0.229\pm 0.002.
\end{equation}
Interestingly, the area of frozen rectangles varies greatly, and a
finite number of rectangles contain a finite fraction of all area.  We
obtain the growth exponent $\alpha$ from extensive numerical
simulations.  Further, we use heuristic arguments to show that this
exponent governs important statistical properties including the final
area distribution and the time-dependent area distribution.

We also study a planar aggregation process where the rectangles are
embedded in a two-dimensional domain and interact only with nearest
neighbors.  We focus on the situation when the domain is an $L\times
L$ square, initially consisting of $N=L^2$ squares of size $1\times
1$.  In each elementary step, an aggregate is chosen at random and in
addition, one of its four sides is also chosen at random. If this side
is shared with a neighboring rectangle, the two rectangles coalesce
into a larger rectangle. The system eventually reaches a jammed
configuration where no two {\it neighboring} rectangles have a common
side. In the planar case, the average number of frozen rectangles is
proportional to system size.  Moreover, frozen rectangles generate a
special tiling of the two-dimensional domain, with the constraint that
two neighboring rectangles may not have a matching side.

\section{The Jammed State}

In our aggregation process, each rectangular cluster has horizontal
size $i$, vertical size $j$, and hence, area $a=i\,j$.  In each
aggregation attempt, we choose two rectangles at random, and
additionally, we randomly choose a side, either horizontal or
vertical.  If the two respective sides are equal in size, the two
rectangles coalesce; otherwise, the rectangles are left intact.
Thus, there are two equivalent aggregation ``channels'' (see figure
\ref{fig-agg})
\begin{subequations}
\begin{align}
(i_1,j)+(i_2,j) &\to (i_1+i_2,j),
\label{process-ver}\\
(i,j_1)+(i,j_2) &\to (i,j_1+j_2),
\label{process-hor}
\end{align}
\end{subequations}
where the first line corresponds to the vertical side, and the second
line, to the horizontal side.

Initially, the system consists of $N$ identical square tiles with
$a=i=j=1$, and consequently, the horizontal size $i$ and the vertical
size $j$ are integers.  Each successful aggregation event reduces the
number of rectangles by one, $n\to n-1$, where $n$ is the total number
of remaining rectangles.  First, we study the mean-field version where
all rectangles interact at a uniform, size-independent, rate. Each
rectangle attempts one horizontal and one vertical aggregation event
per unit time, and hence, time is augmented by the increment $\Delta
t=N/[n(n-1)]$ after each aggregation attempt, $t\to t+\Delta t$.

The process \eqref{process-ver} requires two rectangles with the same
vertical size, and similarly, the process \eqref{process-hor} requires
two rectangles with the same horizontal size. Therefore, aggregation
comes to a halt when each cluster has two unique sizes. In general,
the system evolves toward a ``jammed'' state where $f$ rectangles have
$f$ distinct horizontal sizes and $f$ distinct vertical sizes.

\begin{figure}[t]
\includegraphics[width=0.45\textwidth]{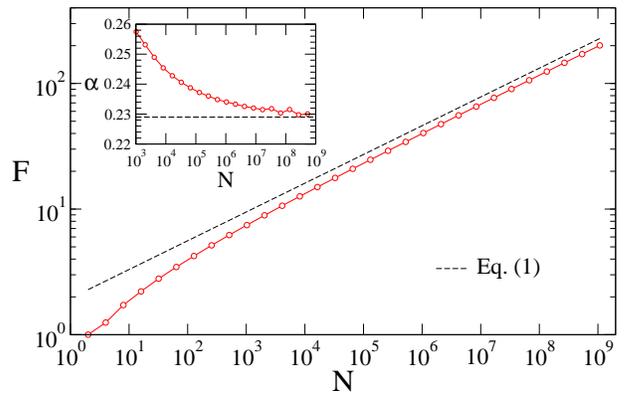}
\caption{The average number of frozen aggregates $F$ versus system
  size $N$.  The inset shows the growth exponent
  \hbox{$\alpha(N)\equiv d\ln F/d\ln N$} versus $N$. The limiting
  value \hbox{$\alpha\equiv \lim_{N\to \infty}\alpha(N)$} quoted in
  Eq.~\eqref{FN} was obtained by noting that $\alpha(N)$ is linear
  $1/F$ when $N$ is sufficiently large.}
\label{fig-FN}
\end{figure}

Our simulations show that the number of aggregates in the jammed state
is a fluctuating quantity. As shown in figure \ref{fig-FN}, the
expected number of frozen rectangles, $F=\langle f\rangle$, grows
algebraically with system size as announced in Eq.~\eqref{FN}. Thus,
the number of frozen clusters grows sub-linearly with system size. We
also measured the variance, $\sigma^2=\langle f^2\rangle -\langle
f\rangle^2$, and found that $\sigma^2\sim F\sim N^\alpha$.  This
behavior contrasts with the behavior typically observed in aggregation
processes where the final outcome is deterministic with the entire
system mass condensing into a single aggregate.

The aggregation processes \eqref{process-ver}--\eqref{process-hor} are
symmetric with respect to orientation, and as long as the initial
conditions are also symmetric, we expect that statistical properties
of the horizontal size and of the vertical size are equivalent.  Yet,
as we discuss below, the width $\omega$ and the length $\ell$ defined
as
\begin{equation}
\label{wl-def}
\omega={\rm min}(i,j), \qquad \ell={\rm max}(i,j),
\end{equation}
are not statistically equivalent.  

For sufficiently-large systems, we find that the jammed state contains
two rectangles of width $\omega=1$: one of horizontal size $i=1$, and
one of vertical size $j=1$.  Similarly, there are two rectangles of
width $\omega=2$, two rectangles of width $\omega=3$, and so on. For
instance, we quote the ordered widths of the $f=13$ frozen rectangles
\begin{equation}
\label{seq}
\{1,1,2,2,3,3,4,4,5,5,6,7,9\}\,,
\end{equation}
that emerged in a simulation of a system of size $N=10^4$ where
$F=13.2$ and $\sigma=1.2$. We note that there may be one or two
rectangles of a given width, and that the width sequence may contain
gaps.  We characterize this sequence using the maximal width
$\omega_{\rm max}$ and the maximal width for which there is a pair of
rectangles, $\omega_{\rm pair}$; for the sequence \eqref{seq},
$\omega_{\rm max}=9$ and $\omega_{\rm pair}=5$.  Our simulations
confirm that both averages $\langle \omega_{\rm max}\rangle$ and
$\langle \omega_{\rm pair}\rangle $ follow the growth law \eqref{FN}.
We deduce that the typical width grows as the total number of frozen
aggregates,
\begin{equation}
\label{wn}
w\sim N^\alpha\,.
\end{equation}

\begin{table}[b]
\begin{tabular}{|c|c|c|c|c|c|c|}
\hline
$\omega$&$1$&$2$&$3$&$4$&$5$& $6$\\
\hline
$m_\omega$ & $0.622$ & $0.182$ & $0.0694$ & $0.0365$ & $0.0214$ & $0.0139$ \\
\hline 
$M_\omega$ & $0.622$ & $0.804$ & $0.873$  & $0.910$  & $0.931$  & $0.945$ \\
\hline
\end{tabular}
\caption{The area fraction $m_\omega$ and the cumulative area fraction 
  $M_\omega=\sum_{\nu\leq \omega}m_\nu$ versus the width $\omega$.}
\label{Tab:mw}
\end{table}

A striking feature of the jammed state is that frozen rectangles of
finite width contain a finite fraction of the total area. In
particular, the two rectangles with width $\omega=1$ contain a
fraction $m_1=0.622$ of all area, while the two rectangles with width
$\omega=2$ contain a fraction $m_2=0.182$ of all area.  Generally, the
fraction $m_\omega$ of all area in rectangles of width $\omega$
approaches a constant in the limit $N\to\infty$.  Table~\ref{Tab:mw}
lists $m_\omega$ and its cumulative sum for $1\leq \omega\leq 6$.
Remarkably, rectangles with width $\omega\leq 6$ contain nearly $95\%$
of the total area.  The area fraction is normalized,
\begin{equation}
\label{norm}
\sum_{\omega=1}^\infty m_\omega=1\,.
\end{equation}
We stress that $m_\omega$ may fluctuate when $N$ is finite, but it
becomes deterministic in the limit $N\to\infty$.  Moreover, the area
$A_\omega$ of rectangles with width $\omega$ is a random quantity with
average $\langle A_\omega\rangle \simeq m_\omega N$ and variance
$\langle A_\omega^2\rangle - \langle A_\omega\rangle^2 \sim N$.

\begin{figure}[t]
\includegraphics[width=0.45\textwidth]{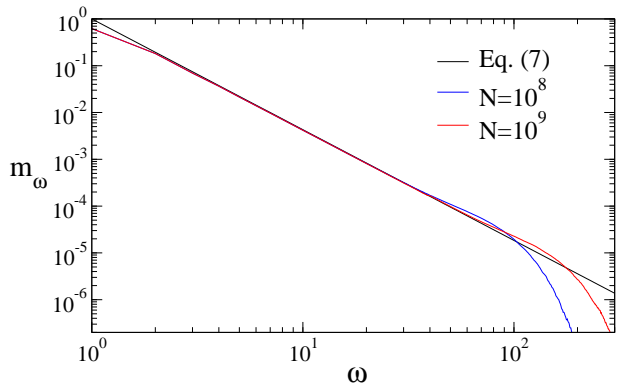}
\caption{The area fraction $m_\omega$ versus width $\omega$, obtained
  from simulations with $N=10^8$ and $N=10^9$.  Also shown For
  reference is the theoretical prediction \eqref{mw-tail} with the
  exponent $\alpha$ from Eq.~\eqref{FN}.}
\label{fig-mw}
\end{figure}   

The length of rectangles of width $\omega$ is proportional to system
size, $\simeq m_\omega N/(2\omega)$, so their aspect ratio scales as
$m_\omega N /(2\omega^2)$. The quantity $m_\omega$ decreases with
width and by definition, the aspect ratio must be larger than
unity. We now postulate that the aspect ratio of the widest rectangles
is of order unity and hence, $m_\omega \sim w^2/N$. By using the
scaling behavior $N\sim w^{1/\alpha}$ which follows from \eqref{wn},
we arrive at our second major result that the area fraction decays
algebraically with width (figure \ref{fig-mw})
\begin{equation}
\label{mw-tail}
m_\omega\sim \omega^{-\gamma}, \qquad \gamma=\alpha^{-1} - 2\,
\end{equation}
when $\omega\gg 1$. Our numerical simulations support this theoretical
prediction. We also note that the normalization \eqref{norm} sets the
bounds $\gamma>1$ and $\alpha<1/3$.

In a finite system, the power-law decay \eqref{mw-tail} holds up to
the scale $w\sim N^\alpha$ which also characterizes the widest
rectangles. Interestingly, rectangles of finite width
\hbox{$\omega={\cal O}(1)$} have an area proportional to system size, 
\hbox{$A\sim N$}, but otherwise, rectangles with the typical width
$w\sim N^\alpha$ have an area that grows sublinearly with system size
\hbox{$A\sim w^2\sim N^{2\alpha}$}. The former rectangles have a large
aspect ratio that is proportional to system size, while the latter
have an aspect ratio of order one. Consequently, a finite number of
clusters have a macroscopic area, while typical clusters contains a
vanishing fraction $N^{2\alpha-1}$ of total area.

\section{Aggregation Kinetics}

We now turn our attention to the time-dependent evolution toward the
jammed state. Let $R_{i,j}(t)$ be the density of rectangles with
horizontal size $i$ and vertical size $j$ at time $t$. This quantity
satisfies the rate equation
\begin{eqnarray}
\label{rate-eq}
\frac{dR_{i,j}}{dt}&=&\sum_{i_1+i_2=i}R_{i_1,j} R_{i_2,j}-2R_{i,j}\sum_{k\geq 1} R_{k,j}\\
&+&\sum_{j_1+j_2=j} R_{i,j_1} R_{i,j_2}-2R_{i,j}\sum_{k\geq 1} R_{i,k},\nonumber 
\end{eqnarray}
and is subject the initial condition
$R_{i,j}(0)=\delta_{i,1}\delta_{j,1}$.  The evolution equation
\eqref{rate-eq} is a straightforward generalization of the
Smoluchowski equations for coagulation \cite{mvs16,mvs17,fl,krb}. The
first two terms on the right-hand side account for the aggregation
process \eqref{process-ver}, with the first term describing gain of a
single large rectangle and the second term describing loss of two
small rectangles.  Similarly, the last two terms account for the
complementary process \eqref{process-hor}.  The rate equations are
mean-field in nature and reflect that all pairs of rectangles interact
at the same rate, set to unity without loss of generality.  In writing
\eqref{rate-eq}, we assumed that the system is infinite.  The
aggregation process conserves area, and accordingly the rate equations
\eqref{rate-eq} conserve the area density $\sum_{i,j} (ij)
R_{i,j}$. Finally, we expect the size density to be symmetric,
$R_{i,j}=R_{j,i}$, because the rate equations and the initial
conditions are symmetric.

The overall density of rectangles $\rho=\sum_{i,j}R_{i,j}$ decays with time 
according to the rate equation 
\begin{eqnarray}
\label{rho-eq}
\frac{d\rho}{dt}=
-\left(\sum_{i,j,k}R_{i,j} R_{k,j}
+\sum_{i,j,k} R_{i,j}R_{i,k}\right)\,.
\end{eqnarray}
This equation is obtained by summing \eqref{rate-eq} over both
indices, and it reflects that in each aggregation event, the total
number of aggregates decreases by one. The two loss terms account for
the two aggregation processes in
\eqref{process-ver}--\eqref{process-hor}. Unlike ordinary aggregation,
the density does not obey a closed equation; more generally, moments
of the size distribution do not obey closed equations.

\begin{figure}[t]
\includegraphics[width=0.45\textwidth]{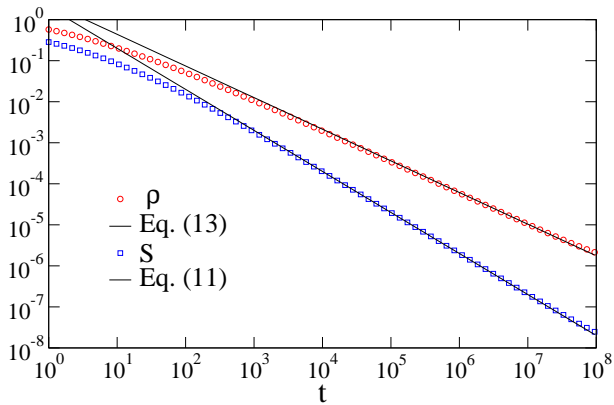}
\caption{The density of aggregates $\rho$ and the density of sticks
  $S$ versus time $t$, obtained from numerical simulations with
  \hbox{$N=10^8$}.  Also shown for reference are the theoretical
  predictions \eqref{S-decay} and \eqref{rho-decay}.}
\label{fig-srho}
\end{figure}

The rate equations \eqref{rate-eq} appear to be analytically
intractable.  Nevertheless, we can gain insights from these equations
by considering the density of ``sticks'', \hbox{$S=\sum_j R_{1,j}$}, namely,
rectangles with minimal horizontal size, $i=1$.  By symmetry, $S$ is
also the density of rectangles with minimal vertical size, \hbox{$S=\sum_i
R_{i,1}$}.  By summing \eqref{rate-eq} over $j$, we find 
\begin{equation}
\label{S-eq}
\frac{dS}{dt} = - S^2 - 2\sum_{i,j}R_{1,j}R_{i,j},
\end{equation}
with $S(0)=1$. The first loss term corresponds to a merger of two sticks
along the vertical side, while the second term corresponds to a merger
of a stick with a rectangle along the horizontal side.  

Based on properties of the jammed state, we argue that the first term
in \eqref{S-eq} is dominant in the long-time limit. Each aggregation
event reduces the total number of aggregates by one. When a finite
system of size $N$ ends up with $f$ frozen aggregates, the total
number of aggregation events equals $N-f$.  Moreover, each aggregation
event generating a stick necessarily involves two sticks. Since frozen
sticks have the macroscopic area $\simeq m_1 N$ with $m_1>1/2$, we
arrive at the remarkable conclusion that in the {\it majority} of all
aggregation events, two sticks combine into a larger stick.
Consequently, the first term in \eqref{S-eq} is dominant, leading to
the closed rate equation $dS/dt =-S^2$ and the power-law decay
(Fig.~\ref{fig-srho})
\begin{equation}
\label{S-decay}
S\simeq t^{-1}\,.
\end{equation}
The simulations confirm that the decay exponent and the prefactor both
equal unity: hence, this behavior is asymptotically exact.

In a finite system, the expected number of sticks with minimal
horizontal size decays as the product of system size and density
\eqref{S-decay}, that is, $\simeq Nt^{-1}$. As discussed in section
II, the jammed state contains two sticks.  Therefore, the jamming time
$\tau$ is simply proportional to system size,
\begin{equation}
\label{tau}
\tau \sim N\,.
\end{equation}

Furthermore, the total density $\rho$ is bounded from below by the
stick density, $\rho\geq S$. In view of the time-dependent behavior
\eqref{S-decay}, we expect that the density decays algebraically with
time, $\rho \sim t^{-\beta}$. In a finite system, the average number
of rectangles scales as \hbox{$N\rho(t)\sim Nt^{-\beta}$}. By
substituting the linear jamming time \eqref{tau} into $N\rho(\tau)$
and matching to the average number of frozen rectangles in \eqref{FN},
we conclude that $\alpha+\beta=1$.  Therefore, the exponent $\alpha$
also governs the decay of the density (Fig.~\ref{fig-srho})
\begin{equation}
\label{rho-decay}
\rho \sim t^{\alpha-1}\,.
\end{equation}

The size distribution of sticks yields additional insights about the
aggregation kinetics.  Both the first loss term in \eqref{S-eq} and
the consequent asymptotic behavior \eqref{S-decay} are precisely the
same as in ordinary aggregation where each aggregate is characterized
solely by its mass \cite{krb}. The size distribution $R_{1,\ell}$ of
sticks with minimal horizontal size and vertical size $\ell$ obeys
\begin{equation}
\label{S1l-eq}
\frac{dR_{1,\ell}}{dt}\! = \!\sum_{i+j=\ell} R_{1,i} R_{1,j} \!- \!2SR_{1,\ell} 
\!-\!2\left(\sum_i R_{i,\ell}\right)R_{1,\ell}\,,
\end{equation}
with the initial condition $R_{1,\ell}(0)=\delta_{1,\ell}$.  The first
two terms describe aggregation events of the type \eqref{process-ver}
that involve two sticks.  Because the asymptotic behavior
\eqref{S-decay} is exact, we expect that theses two terms are
dominant, while the last term, which corresponds to aggregation events 
of the type \eqref{process-hor}, is asymptotically negligible.

\begin{figure}[t]
\includegraphics[width=0.45\textwidth]{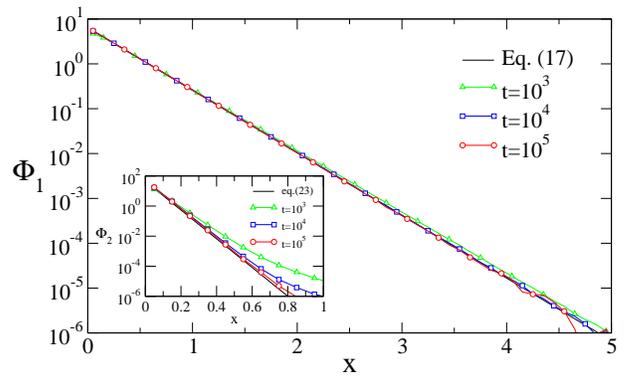}
\caption{The scaling functions $\Phi_1(x)$ and $\Phi_2(x)$ versus the
  scaling variable $x=l/t$.}
\label{fig-phix}
\end{figure}

By dropping the last term in Eq.~\eqref{S1l-eq}, we recover the standard
Smoluchowski rate equation for the size distribution in ordinary
coalescence. In the long-time limit, the size distribution is purely
exponential (Figure \ref{fig-phix})
\begin{equation}
\label{S1l-sol}
S_{1,\ell}\simeq \frac{2}{m_1t^2}\,e^{-2\ell/(m_1\,t)}\,.
\end{equation}
This scaling behavior is consistent with the decay \eqref{S-decay},
and the limiting size of frozen sticks $\sum_\ell \ell S_{1,\ell}\to
m_1/2$ as $t\to\infty$. According to Eq.~\eqref{S1l-sol}, stick length
grows linearly with time, $\ell \sim t$.

Our numerical simulations suggest that rectangles with {\it finite}
width are statistically similar to sticks. Let
$\mathcal{R}_{\omega,l}(t)$ be the density of rectangles with width
$\omega$ and length $\ell$, using the definitions \eqref{wl-def}. This
size distribution adheres to the scaling form (Figure \ref{fig-phix})
\begin{equation}
\label{Phi-def}
\mathcal{R}_{\omega,\ell}(t)\simeq t^{-2}\,\Phi_\omega\big(\ell\, t^{-1}\big)\,,
\end{equation}
as in Eq.~\eqref{S1l-eq}.  The scaling function $\Phi_\omega$
quantifies the density of sticks with length $\ell$ and width
$\omega$. This function is always exponential, although the decay
constant and the normalization depend on width (Figure \ref{fig-phix})
\begin{equation}
\label{Phi-sol}
\Phi_\omega(x)=2\mu_\omega\,e^{-\mu_\omega x}\,, \quad  \mu_\omega=2\omega/m_\omega\,.
\end{equation}
According to equations \eqref{Phi-def}--\eqref{Phi-sol}, rectangles
with width $\omega$ contain a fraction $m_\omega$ of the total area, 
\hbox{$\sum_\ell \omega\ell \mathcal{R}_{\omega,\ell} \to
  m_\omega$} as $t\to \infty$. Moreover, the density $U_\omega
=\sum_\ell \mathcal{R}_{\omega,\ell}$ of rectangles with width $\omega$ and
length $\ell$ decays algebraically with time,
\begin{equation}
\label{Uw-decay}
U_\omega \simeq 2\,t^{-1}\,.
\end{equation}
This behavior can be obtained by repeating the steps leading to
\eqref{S-decay}, which describes the stick density, \hbox{$S\simeq
  U_1/2$}.  The scaling behavior \eqref{Phi-def}--\eqref{Phi-sol} also
shows that the average length of rectangles with width $\omega$ grows
linearly with time, $\langle \ell_\omega\rangle \sim (m_\omega
t)/(2\omega)$, or equivalently, \hbox{$\langle \ell_\omega\rangle
  \simeq v_\omega t$}, with the growth velocity $v_\omega =
m_\omega/(2\omega)$.

\begin{figure}[t]
\includegraphics[width=0.3\textwidth]{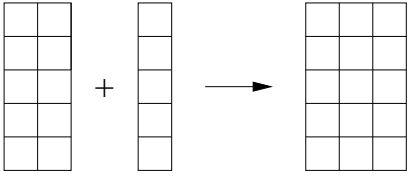}
\caption{Illustration an aggregation process of the type
  \eqref{process-ver} which increases the width.}
\label{fig-agg1}
\end{figure}

Generally, merger events can lead to either wider or longer
rectangles. Figures \ref{fig-agg} and \ref{fig-agg1} both depict
merger events of the type \eqref{process-ver}. In the first scenario,
the width stays the same, but the length grows. However, in the second
scenario, the length stays the same, but width grows.  Thus far, we
have seen that: (i) in a finite system, frozen rectangles with finite
width are macroscopic, (ii) the typical length of rectangles with
finite with grows linearly width time, and (iii) the scaling function
underlying the size distribution is exponential as in ordinary
aggregation. All these results suggest that elongation events are
dominant.

\begin{figure}[t]
\includegraphics[width=0.45\textwidth]{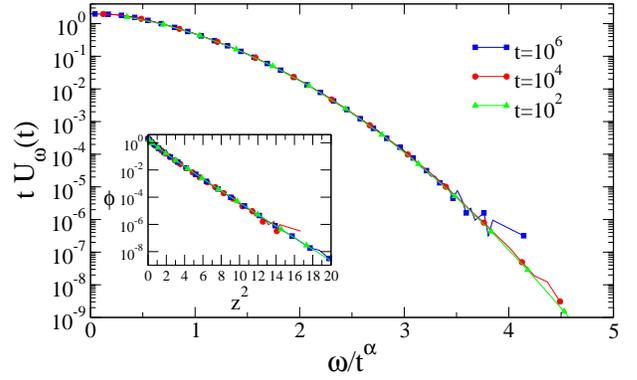}
\caption{The scaled width distribution $t\,U_\omega(t)$ versus the
  scaled width $\omega/t^{\alpha}$. The inset shows the large $z$ tail
  of the scaling function $\phi(z)$.}
\label{fig-Uw}
\end{figure}

The scaling behavior \eqref{Phi-def}--\eqref{Phi-sol} applies for
finite $\omega$ in the long-time limit.  Yet, the typical width does
grow, albeit slowly, with time
\begin{equation}
\label{w-growth}
\omega\sim t^{\alpha}.
\end{equation}
This scale is consistent with the growth law \eqref{FN} and the
jamming time \eqref{tau}, and it can also be obtained by substituting
the density \eqref{rho-decay} and the width \eqref{Uw-decay} into the
estimate $\rho\sim w\,U_w$.  As shown in Figure \ref{fig-Uw}, the
scaled width distribution becomes a universal function of the scaled
width
\begin{equation}
\label{U-scaling}
U_\omega(t)\simeq t^{-1}\phi(\omega t^{-\alpha})\,.
\end{equation}
Equation \eqref{Uw-decay} sets the scaling of the distribution, while the
growth law \eqref{w-growth} sets the scaling of the width.  The
small-$z$ behavior is simply $\phi(0)=2$. Moreover, simulations
suggest that relatively wide rectangles are very rare as the large-$z$
tail is roughly Gaussian, \hbox{$\phi(z)\sim \exp(-{\rm const.}\times
  z^2)$} for $z\gg 1$.

\begin{figure}[t]
\includegraphics[width=0.45\textwidth]{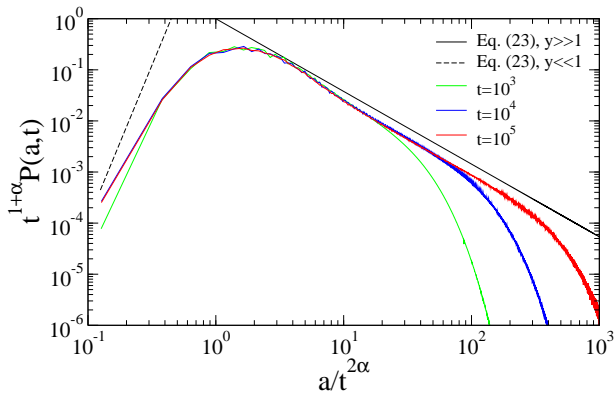}
\caption{The scaled area distribution $t^{1+\alpha}P(a,t)$ versus the
  scaled area $a/t^{2\alpha}$.}
\label{fig-pat}
\end{figure}

Next, we study the density $P(a,t)$ of rectangles with area $a$ at
time $t$.  This quantity satisfies the sum rules
\begin{equation}
\label{pat-norm}
\sum_a P(a,t)=\rho,\qquad \sum_a a P(a,t)=1\,.
\end{equation}
In section II, we noted that while rectangles with finite width have
diverging aspect ratio, rectangles with the typical width \eqref{wn}
have finite aspect ratio. Thus, we expect that the typical width
\eqref{w-growth} governs the typical area, $a\sim w^2 \sim
t^{2\alpha}$. Consequently, the area distribution adheres to the
scaling form (Figure \ref{fig-pat})
\begin{equation}
\label{pat-scaling}
P(a,t)\sim t^{-1-\alpha}\,\varphi\left(a\,t^{-2\alpha}\right)\,.
\end{equation}
Here, the time-dependent prefactor ensures that the total density,
given by the first sum in \eqref{pat-norm}, decays as in
\eqref{rho-decay}.

The scaling function underlying the area distribution, $\varphi(z)$, 
has two algebraic tails 
\begin{equation}
\label{varphi-tails}
\varphi(y)\sim 
\begin{cases}
y^{(3-\alpha)/(2\alpha)}     & y\ll 1, \\
y^{-(1-\alpha)/(1-2\alpha)}   & y\gg 1. 
\end{cases}
\end{equation}
The maximal area grows linearly with time, $a_{\rm max}\sim t$, and
the summation in \eqref{pat-norm} should therefore be carried up to
that scale, $\sum_{a=1}^{t} a\, P(a,t) \sim 1$. The algebraic large-$y$
tail is set by this sum which reflects area conservation.  The minimal
area, $a_{\rm min}=1$, corresponds to tiles. The density of tiles,
$T\equiv R_{1,1}$, decays quite rapidly, $T\sim t^{-4}$, as follows
from $dT/dt =-4ST$ and the stick density \eqref{S-decay}. This rapid
decay sets the remarkably sharp small-$y$ tail in
\eqref{varphi-tails}. The two algebraic tails in \eqref{varphi-tails}
indicate that the area distribution is very broad and that the system
may contain rectangles with areas much smaller than or much larger
than the typical area, $1\ll a \ll t$. In contrast, the length
distribution and the width distribution are much narrower, with
exponential and Gaussian tails, respectively.

The scaling behaviors above show that the exponent $\alpha$ governs
key features of the kinetics including the decay of the density
\eqref{rho-decay}, the typical width \eqref{w-growth}, the width
distribution \eqref{U-scaling}, and the area distribution
\eqref{pat-scaling}. Based on these scaling behaviors we postulate
that the joint size distribution exhibits the scaling behavior
\begin{equation}
\label{Rij-scaling}
R_{i,j}(t)\sim t^{-1-\alpha}G(it^{-\alpha},jt^{-\alpha})\,.
\end{equation}
To comply with the decay law \eqref{rho-decay}, the integral of the
scaling function $\iint d\xi d\eta\, G(\xi,\eta)$ must be
finite. Also, area conservation sets \hbox{$\iint d\xi
  d\eta\,\xi\eta\, G(\xi,\eta)\sim t^{1+\alpha}$} where the upper
limit of integration for both $\xi$ and $\eta$ is of the order
$t^{1-\alpha}$. We did not verify the form \eqref{Rij-scaling}, but
nevertheless, we did confirm one of its implications.  The density of
squares, $H(t)=\sum_i R_{i,i}(t)$, is indeed proportional to the
inverse of time, $H\sim t^{-1}$, as follows from \eqref{Rij-scaling}.
Moreover, the average number of squares in the frozen state is of
order unity as $NH(\tau)\sim 1$. This prediction is consistent with
properties of the jammed state where the widest rectangle has aspect
ratio of order unity, while its area is typically the smallest.

We now examine the evolution of the area fraction $m_\omega(t)$,
defined as the fraction of the total area contained in rectangles with
width $\omega$ at time $t$. This quantity changes only when two
rectangles with the same length merge into a wider rectangle as
illustrated in Figure \ref{fig-agg1}.  The fractions $m_\omega$ evolve
according to
\begin{eqnarray}
\label{mw-eq}
\frac{dm_\omega}{dt} \!=\! 2\!\sum_{i+j=\omega}\!\sum_{\ell} \omega\ell \mathcal{R}_{i,\ell}\mathcal{R}_{j,\ell}
                    \!-\! 4\sum_{j}\sum_{\ell} \omega\ell \mathcal{R}_{j,\ell}\mathcal{R}_{\omega,\ell}\,.
\end{eqnarray}
The first term accounts for gain of area by rectangles of width
$\omega$, and the last term, for loss of area from such rectangles.  One
can check that the rate equation \eqref{mw-eq} conserves the total
area $(d/dt)\sum_\omega m_\omega=0$.

For finite width $\omega$, and in the limit $t\to \infty$, we
substitute the exponential length distribution 
\eqref{Phi-def}-\eqref{Phi-sol} into the rate equation
\eqref{mw-eq}. We then convert sums over $\ell$ into integrals,
$\sum_{\ell\geq 1} \to \int_1^\infty d\ell$. The resulting rate
equation for the area fraction $m_\omega$ satisfies
$dm_\omega/dt\simeq -C_\omega\,t^{-2}$ with the constants
\begin{equation}
\label{cw}
C_\omega=-2\omega\sum_{i+j=\omega}\frac{\mu_i \mu_j}{(\mu_i + \mu_j)^2}
+4\omega\sum_{j}
\frac{\mu_\omega \mu_j}{(\mu_\omega + \mu_j)^2}\,,
\end{equation}
and $\mu_\omega=2\omega/m_\omega(\infty)$.  Therefore, the relaxation
of the area fraction toward the limiting value is similar to
\eqref{Uw-decay},
\begin{equation}
\label{mwt}
m_\omega(t)-m_\omega(\infty)\simeq C_\omega\,t^{-1}\,,
\end{equation}
as $t\to\infty$. Using the fractions $m_\omega$ that were obtained
from numerical simulations (see Table~\ref{Tab:mw} and figure
\eqref{fig-mw}), we have $C_1=1.7$ and $C_2=4.8$. Numerically, we
verified that $m_1(t)$ and $m_2(t)$ evolve according to \eqref{mwt}.

Equation \eqref{mwt} shows that the area fraction decreases
monotonically when $t\to \infty$. Moreover, the growth law
\eqref{w-growth} suggests that there is a time scale $t_\omega\sim
w^{1/\alpha}$ which characterizes the buildup and discharge of area at
width $\omega$.  At early times, $t\ll \omega^{1/\alpha}$, the area
fraction $m_\omega$ increases with time, and at late times, $t\gg
\omega^{1/\alpha}$, the area fraction decreases with time. We thus
anticipate $m_\omega(t_\omega)\propto m_\omega(\infty)$, and using the
tail of the area fraction \eqref{mw-tail}, we find that the constants
in \eqref{cw} are quadratic
\begin{equation}
C_\omega\sim \omega^2\,.
\end{equation} 

Our results show that rectangles grow by two separate processes which
govern the length and the width. In the majority of aggregation
events, the width stays the same but the length grows. This
aggregation process reduces to ordinary one-dimensional aggregation,
allowing us to obtain exact statistical properties of the length which
grows linearly with time.  A minority of aggregation events increase
the width while keeping the length unchanged. This is a much slower
process as the width grows sub-linearly with time, with a rather small
growth exponent.  Obtaining exact statistical properties of the width
and in particular, obtaining the growth exponent $\alpha$ requires a
solution of the full rate equations \eqref{rate-eq} with the
two-variable scaling form \eqref{Rij-scaling}

\section{Planar Aggregation}

We now discuss a complementary aggregation process involving
rectangles that are embedded in two-dimensional space. Initially, the
system consists of $N$ identical tiles with $i=j=1$.  These tiles
form a two-dimensional grid with linear dimension $L$, and hence, area
$N=L^2$. In each successful aggregation event, two {\it neighboring}
rectangles merge and form a larger rectangle. 

\begin{figure}[t]
\includegraphics[width=0.45\textwidth,height=0.45\textwidth]{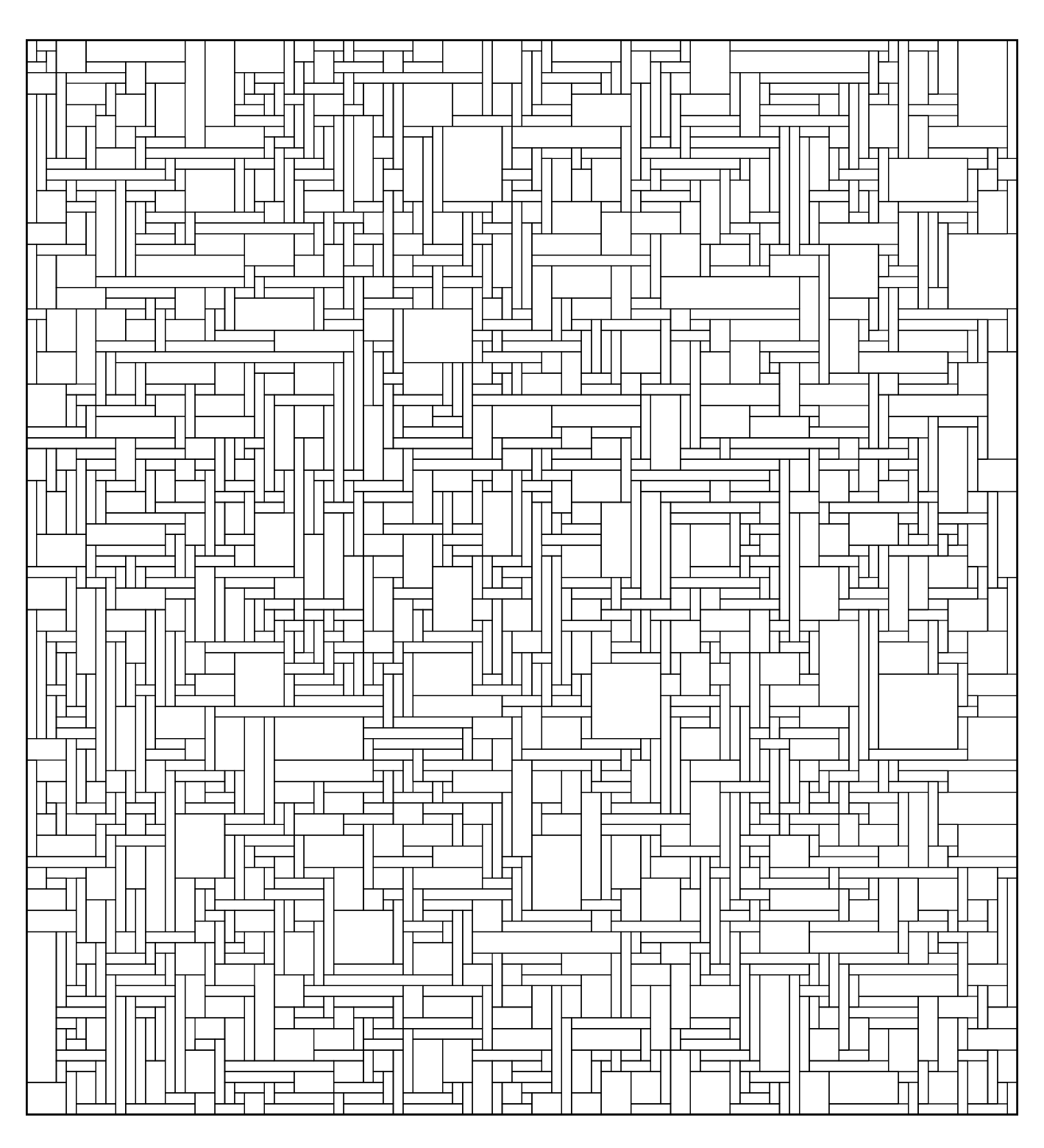}
\caption{A jammed configuration in a system of size $N=10^4$.}
\label{fig-jam}
\end{figure}

There are several ways to realize this planar process. In our
implementation, in each aggregation attempt, one rectangle is chosen
at random, and additionally, one of its four sides is chosen at
random. If this side is shared with a neighboring rectangle, the two
rectangles merge into a larger rectangle as in
\eqref{process-ver} and\eqref{process-hor}.  Time is augmented by the
inverse of the number of rectangles after each aggregation attempt.

In planar aggregation, each rectangle interacts only with nearest
neighbors, in contrast with the mean-field version where each
rectangle interacts with all other rectangles.  Nevertheless, both
systems arrive at a jammed state where aggregation is no longer
possible. In a jammed configuration, two neighboring rectangles never
share a side. Figure \ref{fig-jam} shows a jammed state in a system of
size $N=10^4$.

The jammed state is fascinating as it represents a planar tiling where
the original two-dimensional grid is covered by rectangles. Our
process differs from dimer tiling \cite{pwk,ft,mef,ehl,rjb,cep,ckp} in
that rectangles are heterogeneous and in that there is a constraint
that two neighboring rectangles in the jammed state never share a
side.  Tiling is relevant to thin films growth \cite{sm} and DNA
self-assembly \cite{llry}. We also note that fragmentation of
two-dimensional space, which is the opposite process to the
aggregation considered in this study, also generates tiling by
rectangles \cite{kb94}.

\begin{table}[b]
\begin{tabular}{|c|c|c|c|}
\hline
$\rho_\infty$&$T_\infty$&$S_\infty$&$H_\infty$\\
\hline
$0.1803$ & $9.949\times 10^{-3}$ & $0.1322$ & $2.306\times 10^{-2}$ \\
\hline
\end{tabular}
\caption{The jamming densities of rectangles, tiles, sticks, and
  squares.  These densities were obtained by plotting for example
  $\rho(L)$ versus $L^{-1}$ and using a linear fit.}
\label{Tab-jam}
\end{table}

Qualitatively, the jammed state exhibits orientational ordering as the
orientation of neighboring rectangles are correlated. This correlation
is short-ranged.  Further, rectangles tend to be larger along the
boundary, and they also tend to be aligned with the boundary. For
example, the north and south boundaries have an overpopulation of
horizontal rectangles.

\begin{figure}[t]
\includegraphics[width=0.45\textwidth]{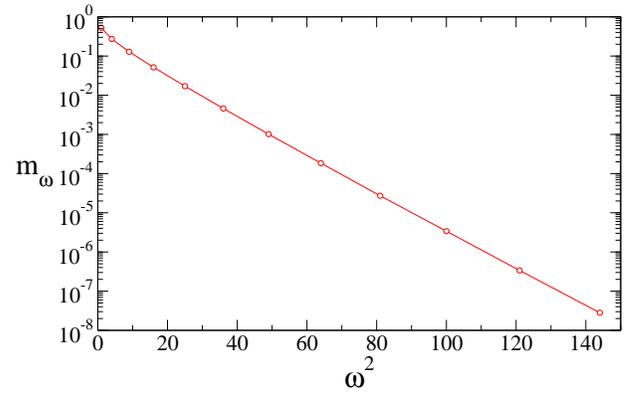}
\caption{The area fraction $m_\omega$ for planar rectangular
  aggregation, from simulations with system size $N=10^8$.}
\label{fig-mw-planar}
\end{figure}

The rectangular tiling can be characterized in several ways.  For
example, the average aspect ratio equals $3.39$.  Further, we measured
the jamming densities of rectangles $\rho_\infty$, tiles $T_\infty$,
sticks $S_\infty$, and squares $H_\infty$ (see Table II). In measuring
these densities, we noticed that finite-size corrections are inversely
proportional to the linear dimension $L$, $\rho(L)-\rho_\infty\sim
L^{-1}$. This behavior suggests that boundary effects extend only a
finite distance from the boundary, and also, that the correlation
length characterizing the orientational order is finite. The finite
density of jammed rectangles, $\rho_\infty=0.1803$, indicates that the
average number of frozen rectangles grows linearly with system size,
\begin{equation}
F\sim N\,.
\end{equation}
This contrasts with the mean-field case where the growth is sub-linear.

We also note that the vast majority of intersections involve three
rectangles, although intersections may rarely be formed by four
rectangles. In the initial state, the tiles form a perfect grid with
each lattice vertex surrounded by four tiles.  We measured $v_n$, the
density of vertices that are surrounded by $n$ rectangles in the
jammed state, with $\sum_{n=1}^4v_n=1$. Internal vertices are
surrounded by a single rectangle, while vertices surrounded by $n=3$
or $n=4$ rectangles correspond to intersections. The vertex densities,
listed in table III, show that a fraction $v_3/(v_3+v_4)=0.988$ of all
intersections are formed by three rectangles, while the complementary
fraction $v_4/(v_3+v_4)=0.012$ of all intersections are formed by four
rectangles.

\begin{table}[b]
\begin{tabular}{|c|c|c|c|c|}
\hline
$n$ & $1$&$2$&$3$&$4$\\
\hline
$v_n$ & $0.215$ & $0.429$ & $0.352$ & $4.26 \times 10^{-3}$ 
\\
\hline
\end{tabular}
\caption{The vertex densities $v_n$.}
\label{Tab-vn}
\end{table}

We also measured the quantity $m_\omega$, the fraction of the total
area in rectangles of width $\omega$.  Figure \ref{fig-mw-planar}
shows that the area fraction has a sharp, Gaussian, tail 
\begin{equation}
\label{mw-tail-planar}
m_\omega\sim \exp\left(-{\rm const.}\times\,\omega^2\right), 
\end{equation}
for $\omega\gg 1$. This decay is much sharper compared with the
power-law behavior \eqref{mw-tail} in the mean-field case.
Consequently, for $\omega\leq 5$, the quantity $m_\omega$ is larger in
the planar case (Table IV).

\begin{figure}[t]
\includegraphics[width=0.45\textwidth]{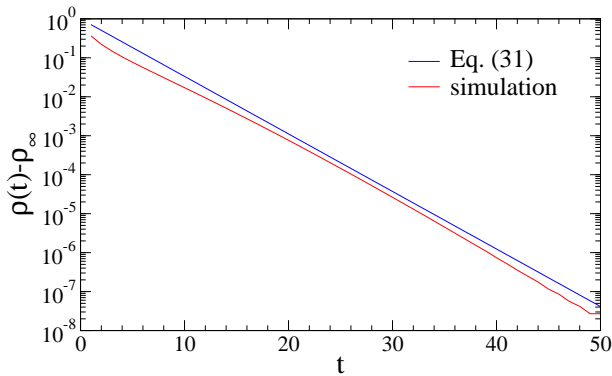}
\caption{The time-dependent density $\rho(t)-\rho_\infty$ versus time.}
\label{fig-ntdif}
\end{figure}

To characterize the approach toward the jammed configuration, we
measured the time-dependent density, and find that the relaxation is
exponential (figure \ref{fig-ntdif})
\begin{equation}
\label{nt-planar}
\rho(t)-\rho_\infty\sim e^{-c\,t}\,.
\end{equation}
with $c=0.34$. This approach is much faster than the algebraic decay
\eqref{rho-decay} that occurs when all pairs of rectangles may
interact.

\begin{table}[b]
\begin{tabular}{|c|c|c|c|c|c|c|}
\hline
$\omega$&$1$&$2$&$3$&$4$&$5$& $6$\\
\hline
$m_\omega$ & $0.525$ & $0.273$ & $0.128$ & $0.0516$ & $0.0170$ & $0.00458$ \\
\hline
\end{tabular}
\caption{The fraction of the total area $m_\omega$ contained in frozen
  rectangles of width $\omega$.}
\end{table}
 
Aggregation of rectangles embedded in a plane has very different
features compared with the mean-field case. The relaxation toward the
jammed state is very fast, jammed rectangles have size and area of the
order unity, and the number of frozen rectangles is proportional to
system size.

\section{Conclusions}

We studied an aggregation process where rectangles with compatible
sides combine to form larger rectangles. This process results in a
jammed state where each rectangle has two unique
dimensions. Aggregation of rectangles exhibits many interesting
features. Some properties of the final state are deterministic while
others are random. For example, the fraction of area in rectangle of
finite width becomes a deterministic quantity in the long-time limit.
However, the number of rectangles in the jammed state is random and it
fluctuates from one realization to another.

While the aggregation process is symmetric with respect to the
horizontal and the vertical sides, there are two different mechanism
that control the growth of the width and the length. In the majority
of aggregation events, two rectangles with the same width combine and
form a longer rectangle with a larger aspect ratio. There are also
rare events where two aggregates of the same length combine to form a
wider rectangle with smaller aspect ratio. Moreover, two distinct
scaling laws characterize the growth of the width and the length.

The growth exponent $\alpha$, which characterizes the growth of the
number of frozen aggregates with system size also governs scaling
properties of the time-dependent behavior. In particular, the width
distribution and the area distribution both adhere to scaling forms in
the long-time limit, and the exponent $\alpha$ characterizes both of
these scaling behaviors. Using heuristic scaling arguments, we related
all power-law exponents that arise in the system to the growth
exponent $\alpha$. Aggregation of rectangles is a special case of
multi-component aggregation as each cluster is characterized by two
sizes. Two distinct growth exponents can emerge in multi-component
aggregation processes, yet, when there are as many conservation laws
as there are components, the scaling exponents can be obtained using
dimensional analysis \cite{kb96,vz,lbd,fg,mm}. The aggregation process
in this study has a single conservation law, and as a result, the
growth exponent $\alpha$ does not follow from heuristic scaling
arguments.

We presented results for a specific realization of the aggregation
process for which the rate equation \eqref{rate-eq} is exact. Yet, we
verified that other variants of this process exhibit a similar
behavior. First, we checked that the growth exponent $\alpha$ in
Eq.~\eqref{FN} is robust even when the vertical channel
\eqref{process-ver} and the horizontal channel \eqref{process-hor} are
realized with different rates. Second, we considered the situation
where aggregates can rotate in order to maximize the fraction of
successful aggregation attempts.  Again, we found that the scaling
properties are robust, although quantities such as the area fraction
are sensitive to the details of the merger process.

We also explored aggregation of $d$-dimensional hyper-rectangles or
``boxes.'' In general, we require that two boxes have a matching
($d-1$ dimensional) facet, such that two boxes combine along this
matching facet to form a larger box. Generally, the system reaches a
jammed state where the expected number of rectangles grows
algebraically with system size as in \eqref{FN}.  The growth exponent
$\alpha$ increases with the spatial dimension: $\alpha=0$, $0.23$,
$0.33$, and $0.40$, for $d=1$, $2$, $3$, and $4$.  The aggregation
process becomes less effective as the dimension increases and
$\alpha\to 1$ when $d\to\infty$.

We also studied a planar realization where only neighboring rectangles
may interact. In this case, the system reaches a jammed state that is
a special rectangular tiling of the plane where two neighboring
rectangles never share a matching side. Statistical properties of the
jammed state are quite different compared with the case where all
rectangles interact. In particular, the number of frozen rectangles is
proportional to system size, and the relaxation toward this state is
exponentially fast. An interesting open question is the entropy of the
jammed configuration. We expect the number of jammed configurations to
grow exponentially with system size, and it would also be interesting
to find out if the entropy of the jammed state depends on the shape of
domain, as is the case for dimer tiling \cite{jb,nd}.

\end{document}